\RequirePackage[l2tabu, orthodox]{nag}
\documentclass[aps, prl, reprint]{revtex4-1}
\usepackage{amsmath,amssymb}
\usepackage{graphicx}
\usepackage{tikz}
\usepackage{color}

\begin{document}
\title{Stabilization and Destabilization of Semiflexible Polymers Induced by Spherical Confinement}
 \author{Martin Marenz}
 \author{Johannes Zierenberg}
\email{Current Address: Max Planck Institute for Dynamics and
Self-Organization, Am Fassberg 17, 37077 G{\"o}ttingen, Germany.}
 \author{Wolfhard Janke}
\affiliation{
  Institut f\"ur Theoretische Physik, 
  Universit\"at Leipzig, 
  Postfach 100\,920, 
  04009 Leipzig, 
  Germany
}

%
\begin{abstract}
  Spherical confinement can act either stabilizing or destabilizing on the collapsed state of a semiflexible polymer.
  General free-energy arguments suggest that the order of the unconstrained collapse transition
  is the distinguishing factor:
  First order implies stabilization, second order causes destabilization.
  We confirm this conjecture by
  Monte Carlo simulations of a coarse-grained
  model 
  for semiflexible polymers whose chain stiffness tunes the transition order.
  The resulting physical picture is potentially relevant for other systems under strong confinement such as proteins and DNA.
\end{abstract}

\maketitle

Confinement is an important element in nature's bag of tricks. Its utilization
ranges from chaperone-mediated protein folding~\cite{hartl2002, young2004} and
subsequent modulation of amyloid formation~\cite{hartl2011,arosio2016}, over
DNA packing in a viral capsid and its translocation into a host
cell~\cite{arsuage2002, arsuage2005, forrey2006, virnau, cao2014,
cacciuto2006a, muthukumar2007, hu2008}, to entropic segregation~\cite{jun2006,
ha2015, azari2015}. Recent results include confinement-induced coexistence of
coil and globule domains along a single DNA chain~\cite{sung2016}. 
Moreover, the same trick may be exploited for practical applications, including
confinement-induced miscibility in polymer blends~\cite{zhu1999}, enhanced
dispersion in nanoparticle-polymer blend films~\cite{chandran2014}, filament
growth in flexible confinement~\cite{vetter2014}, for nanodevices and their
fabrication~\cite{wu2004, claessens2006, reisner2012}, and (DNA) nanopore
sequencing~\cite{branton2008}. It is thus desirable to have a systematic
approach to predict or even tailor the behavior of polymers under confinement.

Semiflexible polymers are a generic model class for a systematic study of the
leading-order effect of confinement on structural properties and thermal
responses in polymeric systems in general. For flexible self-avoiding polymers,
theoretical scaling analyses of the free energy revealed that spherical
confinement is qualitatively different from planar or cylindrical
confinement~\cite{cacciuto2006b}. This is confirmed by mean-field calculations
for (semi)flexible polymers, predicting several scaling regimes of the
confinement-induced change in free energy~\cite{sakaue2006,sakaue2007}. For
self-attracting theta polymers, a change in free energy leads to a change in the
collapse transition temperature. 
Since confinement
generically increases the free energy, it seems surprising that computer 
simulation of flexible polymers revealed a destabilizing shift in the 
collapse temperature~\cite{higuchi2011,marenz2012},
whereas simulations of semiflexible polymers revealed a stabilizing
shift~\cite{higuchi2011}. The latter is consistent with the trend in computational
studies of proteins ~\cite{zhou2001, takagi2003, rathore2006, mittal2008, arkin2012}, 
which can be considered as a special (complex) class of rather rigid semiflexible 
polymers.
However, adding solvent may change the stabilizing nature of confinement 
in this case~\cite{lucent2007}.

In this Letter, we show that stabilization or destabilization of polymers,
induced by a steric spherical confinement, is determined by the type of the free
polymer collapse transition, being either first- or second-order. The relation
to the stiffness of self-attracting semiflexible polymers, exhibiting a rich
structural phase space~\cite{seaton2013,marenz2016,zierenberg2016}, is
straightforward: For flexible polymers the collapse transition from an extended
coil to a globule is known to be a continuous (second-order) transition, while
for stiffer polymers the collapse (or folding) transition from rather stiff rods
into bend linear strands, toroids, or hairpin structures, can be thought of as a
discontinuous (first-order) transition~\cite{zierenberg2016}.  In order to
obtain general results, we combine free-energy arguments in the limit of
infinite chain length with numeric results for coarse-grained polymers
representing biologically more realistic finite chains. Altogether, this allows
us to identify the order of the free-polymer collapse transition as the
distinguishing factor for the stabilization or destabilization upon strong
confinement.

To get a deeper understanding of the basic mechanism, we neglect the chemical
details of complex macromolecules and consider a generic semiflexible
homopolymer model that describes an entire class of semiflexible theta polymers. The
homopolymer is modeled as a linear chain of $N$ monomers connected by stiff
bonds of length $\sigma$. Non-bonded monomers interact via the 12-6
Lennard-Jones potential
\begin{equation}
  V_{\mathrm{LJ}}(r_{ij})=4\epsilon\left[(\sigma/r_{ij})^{12}-(\sigma/r_{ij})^{6}\right],
\end{equation}
where $r_{ij}$ is the distance between two monomers, $\epsilon=1$ sets the
temperature scale, and $\sigma=1$ sets the length scale. This models
self-avoidance and short-range attraction leading to a thermally driven collapse
transition. In addition, semiflexibility is modeled by the worm-like chain
motivated bending potential
\begin{equation}
V_{\mathrm{bend}}(\theta_i)=\kappa(1-\cos\theta_i),
\end{equation}
where $\theta_i$ is the angle between adjacent bonds and $\kappa$ is the
stiffness parameter. The polymer is confined to a steric sphere of radius $R$
with non-elastic repulsion at the shell.

Canonical equilibrium estimates are obtained from histogram
reweighting~\cite{HistRew} of replica-exchange Monte Carlo
simulations~\cite{Nemoto1996} in the entire two-dimensional temperature-stiffness
plane~\cite{marenz2016}. Each
parallel instance samples with the Boltzmann weight $\exp(-\beta E_{\rm LJ}
-\beta\kappa E_{\rm bend})$, where $\beta=1/k_{\rm B}T$ is the inverse
temperature with Boltzmann constant $k_{\rm B}$, $E_{\rm
LJ}=\sum_{i=1}^{N-2}\sum_{j=i+2}^{N} V_{\mathrm{LJ}}(r_{ij})$ and $E_{\rm
bend}=\sum_i^{N-2}V_{\mathrm{bend}}(\theta_i)$. In regular intervals replicas 
exchange their conformations according
to the Boltzmann weights (see,
e.g., Ref.~\cite{marenz2016} for details). 
We validated our data for the free case with estimates from
parallelized multicanonical simulations~\cite{zierenberg2013, bergMuca, jankeMuca}.
In the scaling part, we restrict ourselves to the one-dimensional
replica-exchange approach at fixed $\kappa$.
%
A comprehensive overview is obtained for polymers of length $N=28$, where we
sampled the entire semiflexible range $\kappa\in[0,20]$ in the temperature
interval $T \in [0.1,3.0]$ for sphere radii
$R\in\{5,6,8,10,12,15,20\}$. To obtain the scaling exponents, we selected
integer $\kappa$ values and increased the number of sphere radii $R\in[5,120]$
until we obtained a clear result. We repeated this for shorter ($N=14$) and
longer ($N=42$ and for $\kappa=0$ also $N=64$) chains. 
Error estimates are obtained with the Jackknife method~\cite{efron1982}.

\newcommand{\canRgyr}{\left\langle R_g^2\right\rangle}
\begin{figure}
  \centering
  \includegraphics[width=0.8\linewidth]{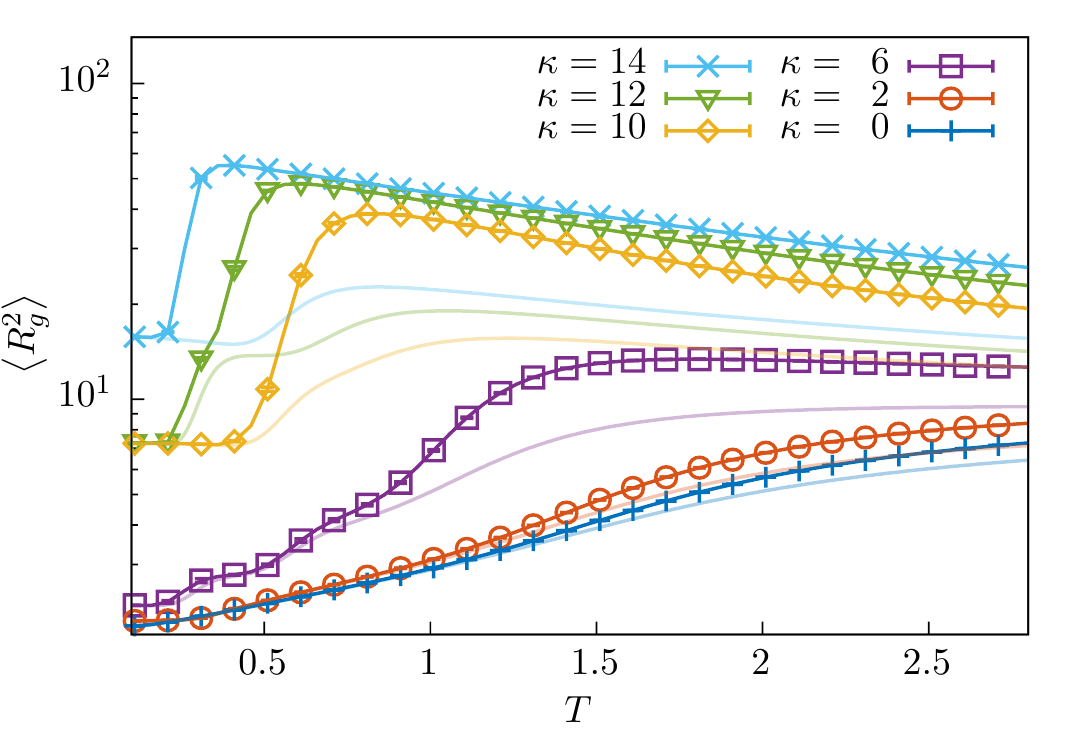}
  \caption{\label{figCanonicalStiffness}%
    Canonical equilibrium extension of a single semiflexible polymer of length
    $N=28$ estimated as the squared radius of gyration $\canRgyr$ for different
    $\kappa$. Intense lines with error bars show the unconfined case and opaque
    lines show the behavior in a sphere of radius $R=8$.
  }
\end{figure}
We begin the discussion of our results by commenting on the effect of stiffness on the
equilibrium collapse transition of a free theta polymer.
Figure~\ref{figCanonicalStiffness} shows a measure of the polymer extension --
the squared radius of gyration $R^2_{g}=\sum_{i=1}^N (r_i-r_{\rm cm})^2/N$,
where $r_{\rm cm}$ is the center of mass. The free case is shown as intense
colored curves, where lines are high-resolution estimates from histogram
reweighting and error bars are shown for some selected points.
For rather flexible polymers, we observe the expected
second-order transition: At high temperatures the polymer is in an extended
conformation (random coil) and, upon temperature decrease, continuously
contracts to a globular equilibrium state. In this regime the bending energy is
a subleading contribution and the problem reduces to the competition of
maximizing conformational entropy by polymer extension and energy minimization
by forming a compact structure.
In contrast, for stiffer polymers we observe a first-order transition: At high
temperatures the polymer is again in an extended conformation (at infinite
temperature again random coil). Upon temperature decrease, the polymer initially
stiffens and extends further towards a stiff rod until a point where it suddenly
folds into more compact conformations such as hairpins or multiple linear
strands~\cite{seaton2013,marenz2016,zierenberg2016}. The sudden decrease hints
towards a (finite-size) ``phase coexistence'', where 
the minimization of Lennard-Jones energy of compact states competes with the
minimization of bending energy of relatively stiff rod-like chain strands.

\begin{figure}
  \centering
  (a)
  \includegraphics[width=0.8\linewidth]{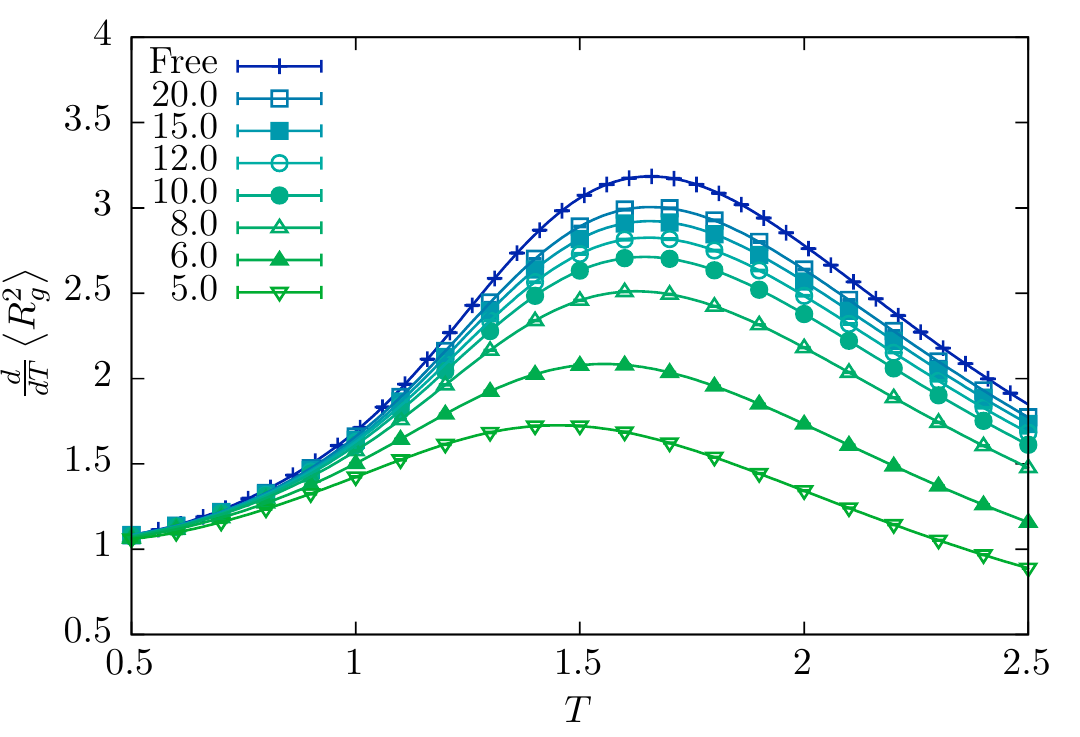}

  (b)
  \includegraphics[width=0.8\linewidth]{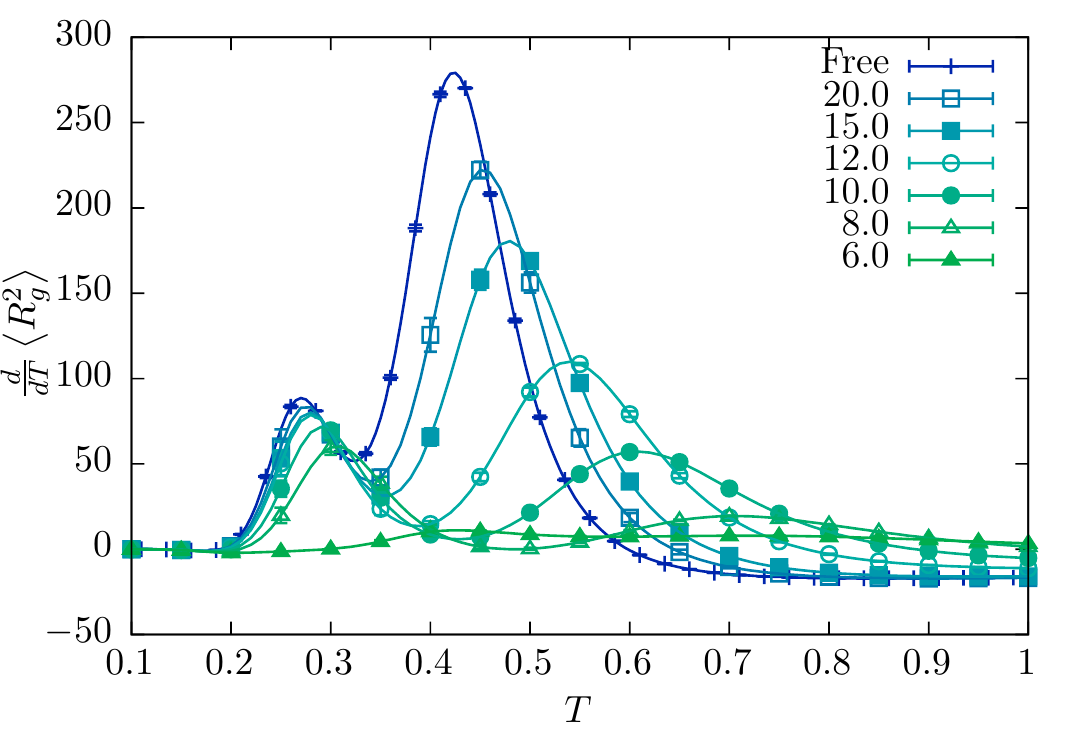}
  \caption{\label{figCanonicalRadius} Canonical thermal derivative of the
    squared radius of gyration $\frac{d}{dT}\left\langle R_g^2 \right\rangle$
    of a single semiflexible polymer ($N=28$) in spherical confinement of different
    radius $R$. Depending on the stiffness, the confinement induces
    destabilization shifting the transition to lower temperatures, e.g., for
    $\kappa=0$ in (a), or induces stabilization shifting the transition to
    higher temperatures, e.g., for $\kappa=12$ in (b).
  }
\end{figure}
The effect of confinement on the canonical estimates of $R^2_g$ can be seen in
Fig.~\ref{figCanonicalStiffness} as opaque lines for the same polymer but in a
steric sphere of small radius. This is here presented for $N=28$ but is
qualitatively similar for shorter ($N=14$) and longer ($N=42$) chains. The
single polymer may no longer be fully extended and we observe a decrease in
$\canRgyr$ at high temperatures for all stiffnesses.
The effect increases with stiffness because the extension of the free polymer
increases with stiffness. As one may expect, the low-temperature behavior is
barely influenced by the confinement.
Depending on the stiffness, the confinement leads to two different modifications
of the collapse transition: 
 
For the case of flexible polymers, the confinement results in a gradual decrease
of the polymer extension along the high-temperature regime which extends over
the transition itself. As a result, the maximal slope here shifts to lower
temperatures.  This is further elucidated for the fully flexible polymer in
Fig.~\ref{figCanonicalRadius}~(a), showing that the peak of the thermal
derivative of $\canRgyr$ shifts to \emph{lower} temperatures with decreasing radius of
the confining sphere.
 
The situation changes for more rigid  polymers, where the spherical confinement
strongly reduces the conformational entropy of only the high-temperature regime
and forces the relatively stiff rods to exhibit a curvature that deviates from
their free counterpart. This appears to be similar to the renormalization of
persistence length known from steric disorder~\cite{schobl}.
Figure~\ref{figCanonicalRadius}~(b) shows the thermal derivative of $\canRgyr$
for a stiffer polymer, demonstrating that a decreasing confining sphere shifts
the peak to \emph{higher} temperatures.
Notice that the lower-temperature peak already shows the next structural
transition (``freezing''), which is only slightly influenced by the spherical
confinement~\cite{marenz2012}.

%
%
\begin{figure}
  \centering
  \includegraphics[width=0.8\linewidth]{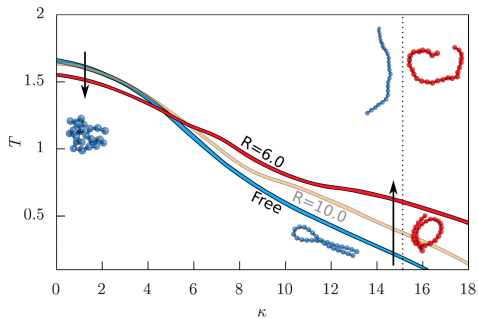}
  \caption{\label{figScheme} Illustration of the stabilizing and destabilizing
    effect induced by spherical confinement on a semiflexible polymer (here
    $N=28$). Colors encode the size of confinement: free (blue), $R=10$
    (orange), and $R=6$ (red). For flexible polymers, the collapse transition
    temperature decreases upon confinement thus destabilizing the collapsed
    state. For stiffer polymers, confinement increases the collapse transition
    temperature thus stabilizing the folded state. This may alter a hairpin
    structure into a purely toroidal state. For intermediate stiffness there is
    a crossover where confinement practically does not alter the transition.
    Exemplary conformations are shown in the color of the respective confinement
    strength at low stiffness ($\kappa=0.3$) and at large stiffness
    ($\kappa=15$, marked by the dashed line).
  }
\end{figure}
The qualitative behavior is summarized 
for the entire range of
semiflexibility in Fig.~\ref{figScheme}.
There is a general trend for flexible polymers to be destabilized by spherical
confinement, i.e., the collapse transition temperature is reduced. 
For stiffer polymers, the transition temperature increases and the folded states
are stabilized at higher temperatures. As for proteins, this may be accompanied
by a change of the structural state, e.g., from hairpin to toroidal structures
as shown in the figure. Similar structures have been observed recently for
double-stranded DNA in confinement~\cite{sung2016}, and as a result of DNA
packing into viral capsids~\cite{arsuage2002, arsuage2005, forrey2006, virnau,
cao2014, footnoteDNA}.
We emphasize that the crossover from destabilization to stabilization ($\kappa
\approx 6$ for $N=28$) indeed roughly coincides with the crossover from a
second-order to a first-order collapse transition in the free semiflexible
polymer~\cite{zierenberg2016}.

\begin{figure}
  \centering
  \includegraphics[width=0.8\linewidth]{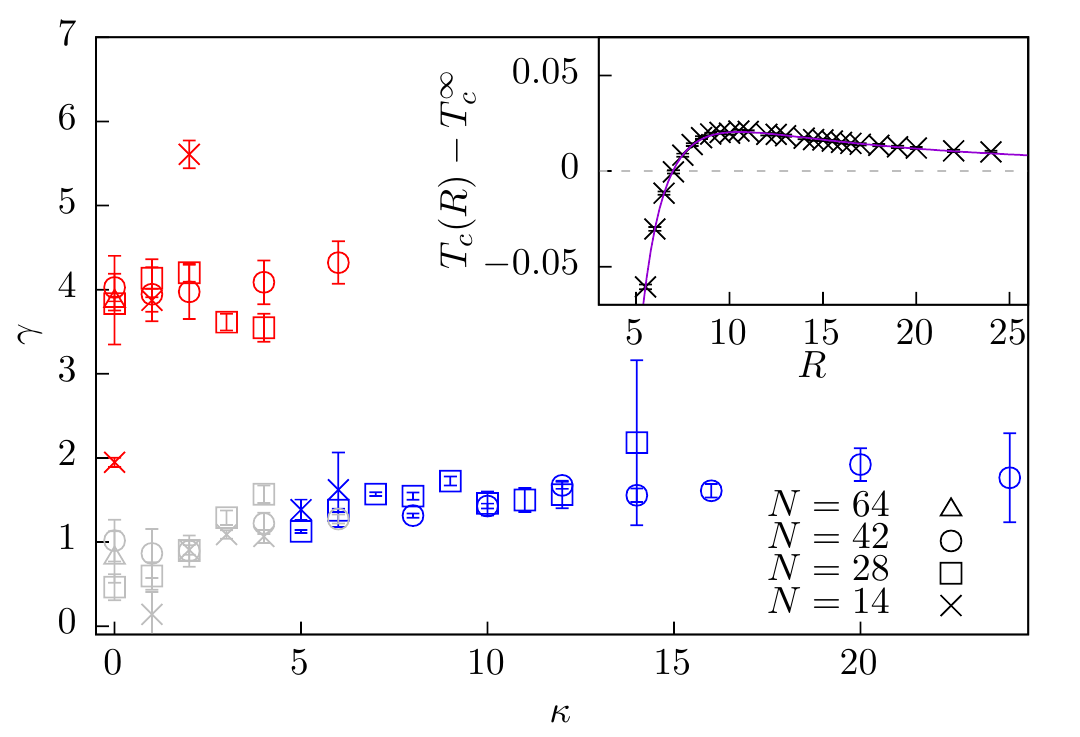}
  \caption{\label{figScaling}%
    Scaling exponent $\gamma$ as a function of stiffness parameter $\kappa$ for
    different $N$. For high $\kappa$, fits to Eq.~\eqref{ansatz1} indicate
    stabilization (blue symbols, $T_c(R)>T_c^{\infty}$) and $\gamma(\kappa)$ shows a
    monotonic behavior. For low $\kappa$, fits to Eq.~\eqref{ansatz2} show both
    subleading stabilizing (gray symbols, $T_c(R)>T_c^{\infty}$) and dominant
    destabilizing (red symbols, $T_c(R)<T_c^{\infty}$) contributions.  
    An example of $T_c(R)-T_c^{\infty}$ in the low-$\kappa$ regime at $\kappa=4$
    is shown in the inset for $N=28$.
  }
\end{figure}

We now want to quantify our observations and provide general free-energy
arguments from which one can deduce the direction and scaling form of the
temperature shift from the type of free-polymer collapse transition.
%
%
Let us begin with the first-order regime of stiffer polymers. At the
transition, the collapsed regime is in coexistence with the extended regime. We
can approximate this in a two-state model, where the system can only change
between a structured state with free energy $F_s$, and an unstructured state
with free energy $F_u$. Coexistence is then expressed in the relation $e^{-\beta
F_s}=e^{-\beta F_u}$ or $0=\beta F_u-\beta F_s$. The spherical confinement
predominantly affects the unstructured regime, decreasing the available entropy
(and even increasing the accessible energy) which increases the free energy $F_u
= F_u^\infty + \Delta F\geq F_u^\infty$, while $F_s\approx F_s^\infty$.  We
consider the temperature-independent correction ansatz $\beta\Delta
F=aR^{-\gamma}$ ($a>0$) in the unstructured (high-temperature) regime. This
ansatz is consistent with all regimes found for semiflexible polymers in
spherical confinement~\cite{sakaue2007} and it yields
\begin{equation}\label{eqTaylorExpansion}
  0 = \beta F_u^\infty - \beta F_s^\infty  + aR^{-\gamma}.
\end{equation}
We can estimate the (inverse) collapse temperature of the confined polymer,
$\beta_c(R)=1/k_B T_c(R)$, by a Taylor expansion around the free polymer
collapse transition, $\beta_c^\infty=1/k_B T_c^{\infty}$. Using
$\partial\beta F/\partial\beta=E$ and $F_s^\infty(\beta_c^\infty)=F_u^\infty(\beta_c^\infty)$ yields 
\begin{equation}
  \beta_c(R) = \beta_c^\infty - aR^{-\gamma}/\Delta E^\infty.
\end{equation}
In general, thermodynamics implies that $\Delta
E^\infty=E^{\infty}_u - E^{\infty}_s\geq0$, where equality may occur for
topological transitions~\cite{marenz2016}, in which case a higher-order
expansion of~(\ref{eqTaylorExpansion}) would be necessary.
For the collapse transition of semiflexible polymers $\Delta E^\infty>0$ and we
obtain a positive temperature shift
\begin{equation}\label{ansatz1}
  T_c(R)-T_c^\infty\propto R^{-\gamma}>0.
\end{equation}
The exponent depends on the free-energy excess that for semiflexible polymers
shows several regimes~\cite{sakaue2007}, some of which are consistent with previous
results for proteins where $\gamma\approx2-3$~\cite{takagi2003}.
Figure~\ref{figScaling} shows that fits of
Eq.~\eqref{ansatz1} to our data for $N=\{14,28,42\}$ indeed yield 
a stabilizing 
shift (blue symbols) 
in the high-$\kappa$ regime
where $\gamma$ is slightly increasing with
$\kappa$ in the range $\gamma\approx 1-2$.

Let us now turn to the second-order regime of flexible polymers. For a
continuous transition, the confinement gradually increases the free energy at
all temperatures, while the effect is of course higher at larger temperatures. A
natural ansatz is to quantify the free-energy increase in terms of the
dimensionless ratio of polymer extension and confinement size $\beta\Delta F\sim
(N^\nu/R)^{x}$, where $\nu$ is the Flory exponent with $\nu\approx3/5$
($\beta<\beta_c$), $\nu=\nu_c=1/2$ ($\beta=\beta_c$), and $\nu=1/3$ ($\beta>\beta_c$)
in three dimensions.  The ratio has to be extensive, i.e., for $N\rightarrow aN$
we expect $F\rightarrow aF$, while $R\rightarrow a^{1/d}R$. It directly follows
for flexible polymers in $d$ dimensions that~\cite{sakaue2006}
\begin{equation}\label{eqDeltaF}
  \beta\Delta
  F \sim \left(\frac{N^{\nu}}{R}\right)^{d/(d\nu-1)}\!\!=N\left(\frac{N}{R^d}\right)^{1/(d\nu-1)}\!\!.
\end{equation}
In order to identify the confinement-induced shift of the collapse transition,
we recall that for polymers of finite length the transition point is signaled by
a shoulder in the specific heat $C_{V}=-k_{\rm
B}\beta^2\left[\frac{\partial^2}{\partial\beta^2}\left(\beta F\right) +
\frac{\partial^2}{\partial\beta^2}\left(\beta\Delta F\right)
\right]$~\cite{vogel2007}. The
obvious temperature-dependent contribution to \eqref{eqDeltaF} is the
step function $\nu(\beta)$, which for finite chains is rounded such that a
linear expansion around the critical point yields
$\nu(\beta)\approx\nu_c-c(\beta-\beta_c)$ with $c$ positive. We thus find that
the confinement-induced change in specific heat scales as {$\Delta
C_{V}\sim-k_{\rm B}\beta^2\left(\frac{N^\nu}{R}\right)^{\frac{d}{(d\nu-1)}}
\left[\frac{d^2c^2}{(d\nu-1)^4}\left(\ln\frac{N}{R^d}\right)^2 -
\frac{2d^2c}{(d\nu-1)^3}\ln\frac{N}{R^d}\right]$}. This change is always
negative and in the limit $R\to\infty$ goes to zero because $R$ grows faster
than $\ln R$. Moreover, the change vanishes for $\beta>\beta_c$ where
$\nu\to1/3$ and is strongest for $\beta<\beta_c$ where $\nu\approx3/5$.
Consequently, with decreasing $R$ the shoulder in the specific heat moves to
larger $\beta$ (smaller $T$) which destabilizes the collapse transition.


In the low-$\kappa$ regime of rather flexible polymers, our numerical results
cannot be described by a single power-law scaling in $R$, see the inset of
Fig.~\ref{figScaling} for a typical example. For large $R$, the confinement
initially weakly stabilizes the collapse transition, while destabilization eventually happens for
smaller $R$. With decreasing $\kappa$, the crossover radius increases and the size
of the temperature shift decreases. Similar observations have been reported for
polymers in slitlike confinement, where narrow slits increase the surface free
energy of the globular state because more monomers are exposed to the
surface~\cite{dai2015}. For the present case, we thus make the extended ansatz  
\begin{equation}\label{ansatz2}
  T_c(R)-T_c^\infty = a_1R^{-\gamma_1} - a_2R^{-\gamma_2}.
\end{equation}
Indeed, the dominant shift in the low-$\kappa$ regime is destabilizing ($a_2\gg
a_1$) with $\gamma_2\approx 4$ (Fig.~\ref{figScaling}, red symbols) in agreement
with our predictions. For the special case of flexible polymers ($\kappa=0$), we
find for small $N$ a purely destabilizing effect ($a_1\approx0$) with
$\gamma_2=1.95(5)$ ($N=14$), while for larger $N$ again a combination of initial
stabilization followed by a clear destabilization with $\gamma_2=3.83(8)$
($N=28$), $\gamma_2=4.0(2)$ ($N=42$), and $\gamma_2=3.9(5)$ ($N=64$). This is
surprisingly consistent with the exponent of the confinement free-energy
change, i.e., $\gamma_2 \approx d/(d\nu-1)\approx15/4=3.75$ (cf.\ Eq.~\eqref{eqDeltaF}) of a flexible
polymer~\cite{sakaue2006,marenz2012}.
It is remarkable that the subdominant stabilization exponent (gray symbols)
seems to extend the increasing trend of $\gamma(\kappa)$ in the high-$\kappa$
regime, which is an interesting starting point for future investigation. 

%
In summary, we have shown both stabilizing and destabilizing effects of
spherical confinement on the entire class of generic semiflexible polymers.
We highlight that using free-energy arguments we have identified the type 
of the free-polymer collapse transition as the distinguishing factor that 
governs the stabilizing (for first-order collapse) and destabilizing (for 
second-order collapse) response on spherical confinement. The type of 
transition in turn can be tuned in our generic model from second- to 
first-order by increasing the chain stiffness~\cite{zierenberg2016}. Our 
result relies on general free-energy arguments and should thus stay valid 
also for more complex polymers, proteins, and DNA. In fact, recent results 
on a special type of flexible polymers with a first-order collapse 
transition show a stabilizing shift in slitlike confinement~\cite{taylor2017}.

Our results thus provide a simple explanation for the diverse results in the
context of polymer collapse and protein folding upon confinement and it would be
worthwile to reassess the so far considered models in the presented framework.
Moreover, our framework may serve as a guide to predict stabilization or
destabilization induced by spherical confinement by studying whether the
transition of a free macromolecule is first- or second-order. In fact, tailoring
the transition behavior, e.g., by DNA origami~\cite{rothemund2006} or by
designing protein structures~\cite{koga2012}, should allow one to attain specific
behavior under confinement. This may find application in targeted drug release
or functionalization of macromolecules in specific confining environments.

\begin{acknowledgments}
We would like to thank Hannes Witt for stimulating discussions. 
This work was in part funded by the European Union and the Free State of Saxony
through the ``S\"achsische AufbauBank'' and by the Deutsche
Forschungsgemeinschaft (DFG) through SFB/TRR 102 (project B04) and Grant No.\
JA~483/31-1.
Additional financial support was obtained by the Deutsch-Franz\"osische
Hochschule (DFH-UFA) through the Doctoral College ``${\mathbb L}^4$'' under
Grant No.\ CDFA-02-07 and the Leipzig Graduate School of Natural Sciences
``BuildMoNa''.
The authors gratefully acknowledge the computing time provided by the John von
Neumann Institute for Computing (NIC) on the supercomputer JURECA at J\"ulich
Supercomputing Centre (JSC) under Grant No.\ HLZ24.
\end{acknowledgments}


\begin{thebibliography}{10}
\bibitem{footnoteDNA}
    More specifically, DNA under typical conditions shows a persistence
    length of $l_p\approx50\text{nm}$. Depending on the salt concentration, the
    thickness $d$ of DNA varies between $2.5\text{nm}$ and $5\text{nm}$ including
    screened electrostatic interactions. This thickness enters in coarse-grained
    models as the bead size $\sigma$. Within the discrete worm-like chain
    approximation we obtain $\kappa/k_BT\approx l_p/\sigma\approx10-20$. Our
    energy scale maps typical temperatures to $k_BT\approx1$, such that
    $\kappa\approx10-20$ depending on the solution~\cite{virnauDNA}. In this
    approximation, chains of length $N=28$ correspond to very short strands of
    about $L=dN\approx(70-140)\text{nm}$ which corresponds to $(210-420)$ base
    pairs.
\end{thebibliography}
\end{document}